\documentclass[preprint,review,doublespacing]{elsart}
\usepackage{graphicx,amssymb,amsmath,times}
\usepackage{subfigure}
\journal{Astroparticle Physics}
\begin{document}
\begin{frontmatter}
\title{Multi-wavelength study of Mrk 421 TeV flare observed with \emph{TACTIC} telescope in February 2010}
\author{K.K. Singh\corauthref{cor}},
\corauth[cor]{Corresponding author.}
\ead{kksastro@barc.gov.in}
\author{K.K. Yadav}, \author{P. Chandra}, \author{S. Sahayanathan}, \author{N. Bhatt}, \author{R.C. Rannot},
\author{A.K. Tickoo}, \author{R. Koul} 
\address {Astrophysical  Sciences  Division, Bhabha Atomic Research Centre. \\
Mumbai - 400 085, India.}
\begin{abstract}
We present results from multi-wavelength study of intense flaring activity from a high 
frequency peaked BL Lac object Mrk 421. The source was observed in its flaring state on 
February 16, 2010 with the $TACTIC$ at energies above 1.5 TeV. Near simultaneous multi-wavelength 
data were obtained from  high energy (MeV-GeV) $\gamma$--ray observations with \emph{Fermi}--LAT, 
X--ray observations by the \emph{Swift} and \emph{MAXI} satellites, optical V-band observation by SPOL 
at \emph{Steward Observatory} and radio 15 GHz observation at OVRO 40 meter-telescope. We have performed 
a detailed spectral and temporal  analysis of $TACTIC$, \emph{Fermi}--LAT  and \emph{Swift}--XRT observations 
of Mrk 421 during February 10--23, 2010 (MJD 55237-55250). The flaring activity of the source is studied 
by investigating the properties of daily light curves from radio to $TeV$ energy range and we present the 
correlation and variability analysis in each energy band. The $TeV$ flare detected by $TACTIC$ on 
February 16, 2010 is well correlated with the activity in lower energy bands. The differential energy spectrum 
of the source, in the energy range 1.5-11 TeV, as measured by $TACTIC$ on this night is described by a power law 
($dN/dE \propto E^{-\Gamma}$) with spectral index $\Gamma = 2.6\pm0.3$. After accounting for absorption of $TeV$  
photons by low energy extragalactic background light photons via pair production, the intrinsic $TeV$ spectrum 
reveals a power law index of $2.3\pm0.3$. Finally the broad band spectral energy distribution of 
the source in flaring state is reproduced using a simple emission model involving synchrotron and 
synchrotron self Compton processes. The obtained parameters are then used to understand the energetics of 
the source during the flaring episode. 
\end{abstract}

\end{frontmatter}
\section{Introduction}
Blazars constitute the most extreme sub-class of active galactic nuclei (AGN). The observed broad-band radiation
from blazars is characterized by non-thermal emission extending from radio to very high energy (VHE) $\gamma$-rays. 
The detection of energetic $\gamma$--rays and rapid variability observed in many blazars suggest, the emission to 
arise from a relativistic jet oriented very close to the line of sight of the observer [1]. Under the current picture 
of AGN, the jet is powered by a super massive black hole at the center of host galaxy and accreting matter from it. 
Blazars are further classified into flat spectrum radio quasars (FSRQ) and BL Lacs depending upon the presence/absence 
of emission/absorption line features. Due to relativistic Doppler boosting, the non thermal emission from the jet 
dominates the entire spectral energy distribution (SED) of blazars. The radiation from blazars, over entire 
electromagnetic spectrum, is extremely variable at different time scales ranging from few minutes to years. 
The variability timescale of blazars can be used to constrain the emission region size through light travel time 
effects and also in understanding the underlying particle acceleration mechanisms [2,3]. 
\par
The broad band SED of blazars consists of two peaks with the first one at IR/Optical/X--ray energies and 
the second at gamma--ray energies [4]. The low energy component is commonly attributed to the synchrotron 
emission from the ultra-relativistic leptons in the jet. On the other hand, the physical mechanisms proposed 
to model the high energy (HE) component include: synchrotron self Compton emission (SSC models) [5], 
Compton up scattering of external photons (EC models) from accretion disk [6], broad line region (BLR) [7], or 
cascades produced by high energy protons [8,9,10]. Therefore, $\gamma$--ray observations of blazars in flaring 
state along with simultaneous multi-wavelength data are important tools to validate the emission models.
\par 
Mrk 421 (z=0.031, 134 Mpc) is the first HBL observed in VHE range [11]. Since its detection, the source has 
been a frequent target for almost all existing ground based imaging atmospheric Cherenkov telescopes 
(IACTs) and also other Multi-Wavelength (MW) campaigns. Recent VHE observations of this source are reported by 
various groups [12,13,14,15,16]. Long term observations of Mrk 421 during 2009--10 with $TACTIC$  have been 
reported in [17]. The source was observed in high flaring state during February 2010 in X--ray and $\gamma$-ray 
bands [18,19,20,21]. The X--ray and HE $\gamma$-ray flares of this source during February 2010 observed by 
\emph{Swift} and \emph{Fermi} respectively have been studied by [22].
\par
Mrk 421 was also observed with $VERITAS$ and $HESS$ telescopes during Februaray 2010. The $VERITAS$ telescope observed an 
unusual bright flare on February 17, 2010 (MJD 55244) in TeV $\gamma$--rays reaching a flux level of approximately 8 Crab 
Units with a variability timescale of few minutes [23]. The flaring episode comprising about 5 hours of observations, 
yields presence of a $\gamma$--ray signal at statistical significance of 256$\sigma$. The spectrum is described by a 
power law with exponential cutoff: $dN/dE=N_0(E/E_0)^{-\Gamma} exp(-E/E_{cut})$,  with normalization 
$N_0=(5.28\pm0.09)\times10^{-10}$ $cm^{-2} s^{-1} TeV^{-1}$, spectral index $\Gamma=1.77\pm0.02$ and an exponential 
cutoff energy $E_{cut}=4.06\pm0.2$ TeV. Triggered by $VERITAS$  high state detection, the source was followed up by 
$HESS$ telescope during February 17-20, 2010 (MJD 55244.96 - 55246.96) at an average zenith angle of $62^{\circ}$ [24].
Analysis of about 5.4 hours of good quality data, obtained after applying various quality checks, yields an excess of 
2112 events at statistical significance of 86.5$\sigma$ with flux level varying from 1.4 to 4.8 Crab Units. The time 
averaged energy spectrum is characterized by a power law with exponential cutoff with normalization $N_0=(1.96\pm0.32)\times10^{-11}$ 
$cm^{-2} s^{-1} TeV^{-1}$, spectral index $\Gamma=2.05\pm0.22$ and an exponential cutoff energy $E_{cut}=3.4\pm0.6$ TeV.  
\par
In the present work, we perform a detailed temporal study of the VHE data collected during February 10--23, 2010 with 
$TACTIC$. We supplement this with near simultaneous MW data in low energy bands to study the bright TeV flare. In addition, 
we also present the results from the spectral study of the flaring data recorded on February 16, 2010 (MJD 55243). 
The organization of this paper is as follows: In Section 2, we briefly describe the $TACTIC$. The MW data set and their 
detailed analysis procedure used in the present study are presented in Section 3. The temporal studies of all data sets are 
reported in Section 4. In Section 5, we describe detailed spectral analysis of the flare on February 16, 2010. The SED modelling 
of the flaring data is presented in Section 6. Finally, in Section 7 we discuss our results and conclusions. Throughout this 
paper we adopt $\Lambda$CDM cosmology with parameters, $H_{0}$=70 km $s^{-1}$ $Mpc^{-1}$, $\Omega_{m}$=0.27 and $\Omega_{\Lambda}$=0.73.

\section{TACTIC Telescope}
The $TACTIC$ \emph{(TeV Atmospheric Cherenkov Telescope with Imaging Camera)} is located at Mount Abu 
( $24.6^{\circ}$ N, $72.7^{\circ}$ E, 1300 m asl), Rajasthan, India [25]. The telescope is equipped 
with a F/1-type tesselated light collector of $\sim$9.5 $m^{2}$ area consisting of 34 front-face aluminium coated, 
spherical glass mirror facets of 60 cm diameter. The facets have been pre-aligned to produce an on-axis spot size of 
$\sim$$0.3^{\circ}$ at the focal plane. The telescope deploys a 349-pixel photo-multiplier tube (ETL 9083UVB) based imaging 
camera, with a uniform pixel resolution of $\sim$$0.3^{\circ}$ and a field of view of $\sim6^{\circ}\times6^{\circ}$ 
to record the fast snapshot of atmospheric Cherenkov events. The data in the present work have been obtained with 
inner 225 pixels and the inner most 121 pixels (11$\times$11 matrix) covering a field of view of 
($\sim$$3.4^{\circ}\times3.4^{\circ}$) were used for generating the event trigger based on nearest neighbour 
non-colinear triplet trigger logic. The triggered events are digitized by CAMAC based 12-bit charge to digital 
converters. The telescope has a sensitivity of detecting Crab nebula at  5$\sigma$ significance level  in  25 hours 
of observation time above 1.5 TeV. Further details regarding the long term performance evaluation of the telescope 
based on 400 hours of data collected on Crab Nebula during 2003-2010 are reported in [26].


\section{Multi-Wavelength observations and Data Analysis}
We use $TACTIC$ data collected from Mrk 421 during its flaring state in February 2010 along with MW archival 
data from \emph{Fermi}--LAT, \emph{Swift}--XRT/BAT, \emph{MAXI}, SPOL and OVRO. Observational details and 
analysis procedure are discussed in the following subsections. 

\subsection{$TACTIC$ observations and TeV-data}
An intense $TeV$ flaring activity of  Mrk 421 was observed with $TACTIC$ during February 2010. After performing data 
quality checks [27], we obtained 48.4 hours of clean data spanning over 12 nights of observation. 
The standard data quality checks for $TACTIC$ involve $(i)$ conformity of the expected prompt 
coincidence rate (PCR) trend with zenith angle, $(ii)$ compatibility of the arrival time of cosmic-ray events 
with the Poissonian statistics and $(iii)$ steady behaviour of chance coincidence rate (CCR) with time. 
The data recorded with $TACTIC$ were analysed using standard Hillas parameterization technique [28]. 
Each Cherenkov image is characterized by various image parameters like LENGTH (L), WIDTH (W), DISTANCE (D), 
ALPHA ($\alpha$), SIZE (S) and FRAC2 (F2) using moment analysis methodology. While the shapes of roughly  
elliptical images are described by L and W parameters, their location and orientation in the 
telescope field of view are specified by D and $\alpha$ parameters respectively. Parameter F2 is defined 
as the ratio of sum of the two highest amplitude pixels to the total image size. The standard dynamic supercut 
procedure [29, 30] is used to segregate $\gamma$- like images from background images of cosmic--rays. 
The $\gamma$--ray selection criteria obtained on the basis of Monte Carlo simulations carried out for $TACTIC$ 
telescope are given in Table 1. The $\gamma$--ray signal is extracted from cosmic--ray background using frequency 
distribution of $\alpha$--parameter after applying the set of image parameter cuts presented in Table 1. The 
distribution of $\alpha$--parameter is expected to be flat for cosmic--ray background due to its isotropic 
behaviour, whereas for $\gamma$--ray events coming from a point source, the distribution is expected to 
show a peak at smaller $\alpha$ values. The present analysis of $TACTIC$ data from Mrk 421 direction during 
February 10--23, 2010 resulted in an excess of 737$\pm$87 $\gamma$--ray like events corresponding to a 
statistical significance of 8.46$\sigma$ in nearly 48 hours. Observation of flaring activity on February 16, 2010 
(MJD 55243) alone yields 172$\pm$30 $\gamma$-like events corresponding to a statistical significance of 
5.92$\sigma$ in $\sim$ 4.9 hours. Fig.1 gives the $\alpha$--distribution of one day data collected during 
the flare on February 16, 2010. The energy of each $\gamma$-like event is reconstructed using an artificial neural 
network (ANN) based methodology on the basis of its zenith angle, SIZE and DISTANCE. The procedure followed by us 
uses 3:30:1 (i.e. 3 nodes in the input layer, 30 nodes in the hidden layer and 1 node in the output layer) configuration 
of the ANN and yields an energy resolution of $\sim$26$\%$ [31].

\begin{table}
\caption{Dynamic Supercuts  selection  criteria used for analyzing the $TACTIC$ data.}
\label{tab:cuts}
\vspace{0.3cm}
\begin{center}
\begin{tabular}{lcc}
\hline
Parameters  	&Cuts Value \\ 
\hline
LENGTH (L)      &0.11$^{\circ}$ $\le$  L $\le$ (0.235 + 0.0265 $\times$ln S)$^{\circ}$ \\
WIDTH (W)       &0.065$^{\circ}$ $\le$ W $\le$ (0.085 + 0.0120 $\times$ln S)$^{\circ}$ \\ 
DISTANCE (D)    &0.5$^{\circ}$ $\le$  D $\le$ (1.27 Cos$^{0.88}$ $\Theta$)$^{\circ}$ ($\Theta$= Zenith angle) \\ 
SIZE (S)        &S $\ge$  485 dc (8.5 digital counts = 1.0pe)\\ 
ALPHA ($\alpha$)&$\alpha$ $\le$ 18$^{\circ}$ \\ 
FRAC2 (F2)      &F2 $\ge$  0.38\\ 
\hline
\end{tabular}
\end{center}
\end{table} 

\begin{figure}
\begin{center}
\includegraphics*[width=1.0\textwidth,angle=0]{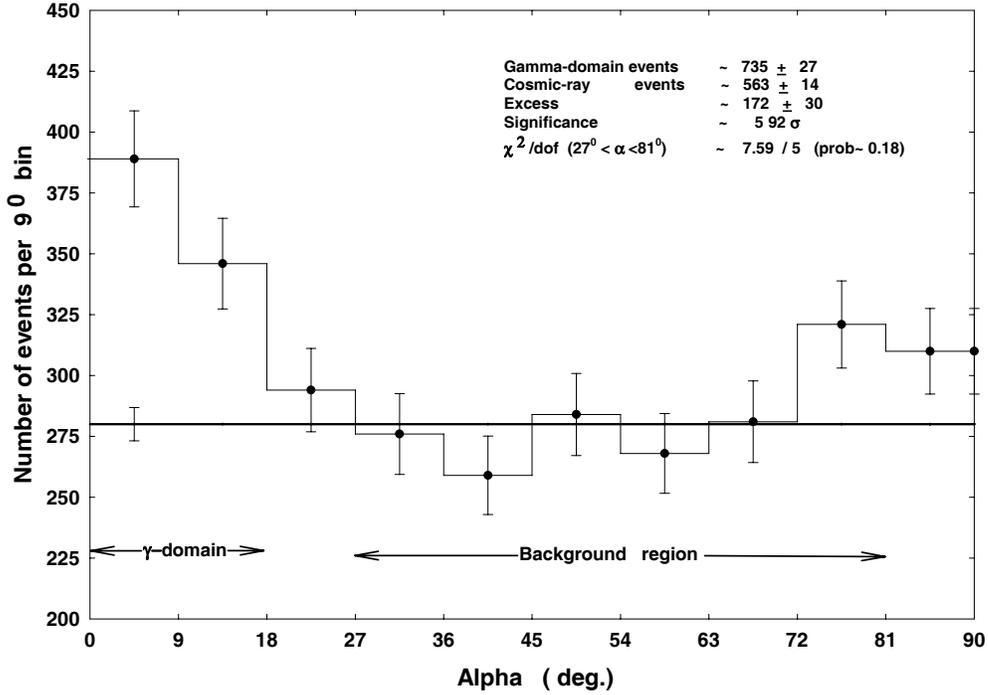}
\caption{On-source alpha distribution of Mrk 421 for $\sim$ 4.9h of data collected during flare on February 16, 2010. 
The horizontal line represents the expected background in $\gamma$-domain.}
\end{center}
\end{figure}
\subsection{\textit{Fermi}--LAT data}
The \textit{Fermi}--LAT \emph{(Large Area Telescope)} is a pair production ($e^{-}$$e^{+}$) telescope sensitive to $\gamma$--ray 
photons in the  energy range 20 MeV--300 GeV [32]. The telescope performance is characterized by a typical energy 
resolution of $\sim$10$\%$ and angular resolution better than $1^{\circ}$ at energies above 1 GeV. With a large 
field of view $\ge$2.4 sr, LAT observes the entire sky every 3 hours in survey mode. The LAT data used in 
this work were collected from MJD 55237 (February 10, 2010) to MJD 55250 (February 23, 2010), the period which 
overlaps with the $\gamma$--ray flare of Mrk 421. The data were obtained from FSSC 
archive\footnote{{http://fermi.gsfc.nasa.gov/ssc/data/access}} and the data 
analysis was performed using the standard \textit{Fermi} Science-Tools software package (version v9r27p1). Only diffuse 
class events in the energy range 100 MeV--100 GeV from a circular region of interest (ROI) with radius $15^{\circ}$  were 
included in the analysis. The set of instrument response functions \emph{P7SOURCE$\_$V6} were used. Events with zenith angle 
$>$ $105^{\circ}$ were filtered out to avoid the Earth albedo as suggested by \textit{Fermi}-LAT team. The data 
obtained in this manner were analyzed using unbinned maximum likelihood algorithm implemented in \textit{gtlike} tool
 which is also a part of  Science-Tools package. The background model used to extract $\gamma$--ray signal from the source
 includes two components: galactic diffuse emission and isotropic background emission. The galactic diffuse component is 
parameterized by the map cube file \textit{gal$\_$2yearp7v6$\_$v0.fits}. The isotropic background component (sum of residual 
instrumental background and extragalactic diffuse $\gamma$--ray background) was included using the standard model file
 \textit{iso$\_$p7v6source.txt}. The spectra of  Mrk 421 and other point sources in ROI were fitted with a power law defined
 as:
\begin{equation}
N(E) = \left[\frac{(\Gamma+1)N_o}{(E_{max}^{\Gamma+1}-E_{min}^{\Gamma+1})}\right]\,E^\Gamma  
\end{equation}
where $\Gamma$ is the photon index, $N_{0}$ is normalization constant, $E_{min}$ and $E_{max}$ are the lower and upper limits
 of the energy interval selected for Likelihood analysis. We derived the daily light curve and differential photon spectrum of
 Mrk 421 in the energy range 100 MeV--100 GeV using the methodology described above. 

\subsection{X--ray data}
X--ray data were obtained with the telescopes on \textit{Swift} [33] and \textit{MAXI} [34] satellites. The \textit{Swift}
satellite is equipped with three telescopes: BAT \emph{(Burst Alert Telescope)} covering the energy range 15--150 keV [35], 
XRT \emph{(X--Ray Telescope)} covering 0.2--10 keV energy band [36], and UVOT \emph{(Ultra-Violet/Optical telescope)} over 
the wavelength range 180--600 nm [37]. We used \textit{Swift}--XRT archival data from MJD 55237 (February 10, 2010) to 
MJD 55250 (February 23, 2010) which were analyzed using \textit{xrtpipeline} utility available with the HEASoft package. 
We produced the light curves and spectra for each day using spectral analysis package XSPEC. 
A daily average flux in the energy range 15--50 keV detected by \textit{Swift}--BAT was obtained from online  data archive 
\footnote{{http://heasarc.nasa.gov/docs/swift/results/transients}}. We also used X--ray data in the energy range 2--20 keV 
observed by X--ray instrument onboard \textit{MAXI} satellite from the website\footnote{{http://maxi.riken.jp/top/index.php}}.

\subsection{Optical and Radio data}
\textit{Fermi} MW observing support program provides data publicly for regular or targeted observation of 
blazars\footnote{{http://fermi.gsfc.nasa.gov/ssc/observations/multi/programs}}. We obtained the V--band optical 
data of Mrk 421 from the SPOL CCD Imaging/Spectropolarimeter [38] at Steward Observatory, University of 
Arizona\footnote{{http://james.as.arizona.edu/~psmith/Fermi/}}. The 15 GHz radio data were obtained from 40m Owens Valley 
Radio Observatory (OVRO) [39]\footnote{{http://www.astro.caltech.edu/ovroblazars/data}}. 


\section{Temporal study}
\subsection{Light curve analysis}
The MW light curves of Mrk 421 during February 10--23, 2010 (MJD 55237--55250) in TeV, MeV-GeV, X--ray, optical and 
radio bands are shown in Fig. 2 (a-g). The source was observed in a high state in $\gamma$--ray and X--ray energy bands 
on February 16, 2010 (MJD 55243). Fig. 2(a) shows the daily averaged light curve of TeV photons detected with the 
$TACTIC$. We observe that the TeV $\gamma$--ray flux starts increasing from February 15, 2010 (MJD 55242), 
attains a peak on February 16, 2010 (MJD 55243) and finally decays gradually. Fig. 2(b-g) correspond to observations 
with \emph{Fermi}--LAT , \emph{Swift}--BAT, MAXI, \emph{Swift}--XRT, SPOL and OVRO respectively. From the figure it is 
apparent that the TeV flaring activity detected by $TACTIC$ is accompanied by enhanced activity in lower energy bands 
from MeV-GeV to X--rays and a mild change in V-magnitude. The radio observations are available only for five days and no 
significant flux variations are observed during this period. If we attribute the observed flare to jet activity, 
non-detection of radio variation can be associated with the significant absorption at these energies due to synchrotron 
self absorption process. The MW flaring on February 16, 2010 (MJD 55243) is followed by an enhanced activity in X--rays on 
February 22, 2010 (MJD 55249). Thus X--ray data show two consecutive flares during the period February 10--23, 2010. 
However, because of the absence of TeV data on February 22, 2010, we concentrate only on the flare detected in MW  regime on 
February 16, 2010 (MJD 55243) in the present study. The light curves in all energy bands (radio to TeV) shown in Fig. 2 are 
fitted with a steady emission during  February 10--23, 2010 to identify the corresponding flare on February 16, 2010 (MJD 55243) 
and the results are summarized in Table 2. We also use e-folding time scale method [40] to compute the variability time 
scale ($t_{var}$) for each light curve. This method allows computation of variability time using just two flux measurements 
and it does not require any fitting or minimization procedure. From the light curve analysis, we observe that Mrk 421 shows a 
time variable broad band emission from X--rays to TeV $\gamma$--rays with a temporal variability of approximately one day. 
\begin{table}
\caption{Summary of MW light curves of Mrk 421 during February 10-23, 2010 (MJD 55237--55250).}
\vspace{0.3cm}
\begin{center}
\begin{tabular}{lclclclc}
\hline
Instrument &Energy Band	&Constant emission    &Flare Emission (16 Feb. 2010)\\ 
\hline
TACTIC  &1.5-11 TeV &(1.0$\pm$0.2)$\times$10$^{-11}$ ph $cm^{-2} s^{-1}$ &(2.5$\pm$0.4)$\times$10$^{-11}$ ph $cm^{-2} s^{-1}$\\ 
LAT	&0.1-100 GeV&(7.2$\pm$0.3)$\times$10$^{-8}$ ph $cm^{-2} s^{-1}$&(3.2$\pm$0.9)$\times$10$^{-7}$ ph $cm^{-2} s^{-1}$\\
BAT	&15-50 keV  &(5.9$\pm$0.1)$\times$10$^{-3}$ cts $cm^{-2} s^{-1}$&(15.7$\pm$0.8)$\times$10$^{-3}$ cts $cm^{-2} s^{-1}$\\
MAXI	&2-20 keV   &(18.0$\pm$2.4)$\times$10$^{-2}$ ph $cm^{-2} s^{-1}$&(46.0$\pm$4.0)$\times$10$^{-2}$ ph $cm^{-2} s^{-1}$\\
XRT	&0.2-10 keV &(1.6$\pm$0.3)$\times$10$^{-9}$ erg $cm^{-2} s^{-1}$&(2.7$\pm$0.6)$\times$10$^{-9}$ erg $cm^{-2} s^{-1}$\\	
SPOL	&V-band (Optical)&(13.09$\pm$0.02) mag.				&(13.04$\pm$0.02) mag.\\
OVRO	&15 GHz (Radio)&(0.445$\pm$0.004) Jy   				&--\\
\hline
\end{tabular}
\end{center}
\end{table}
\begin{figure}
\begin{center}
\includegraphics*[width=1.0\textwidth]{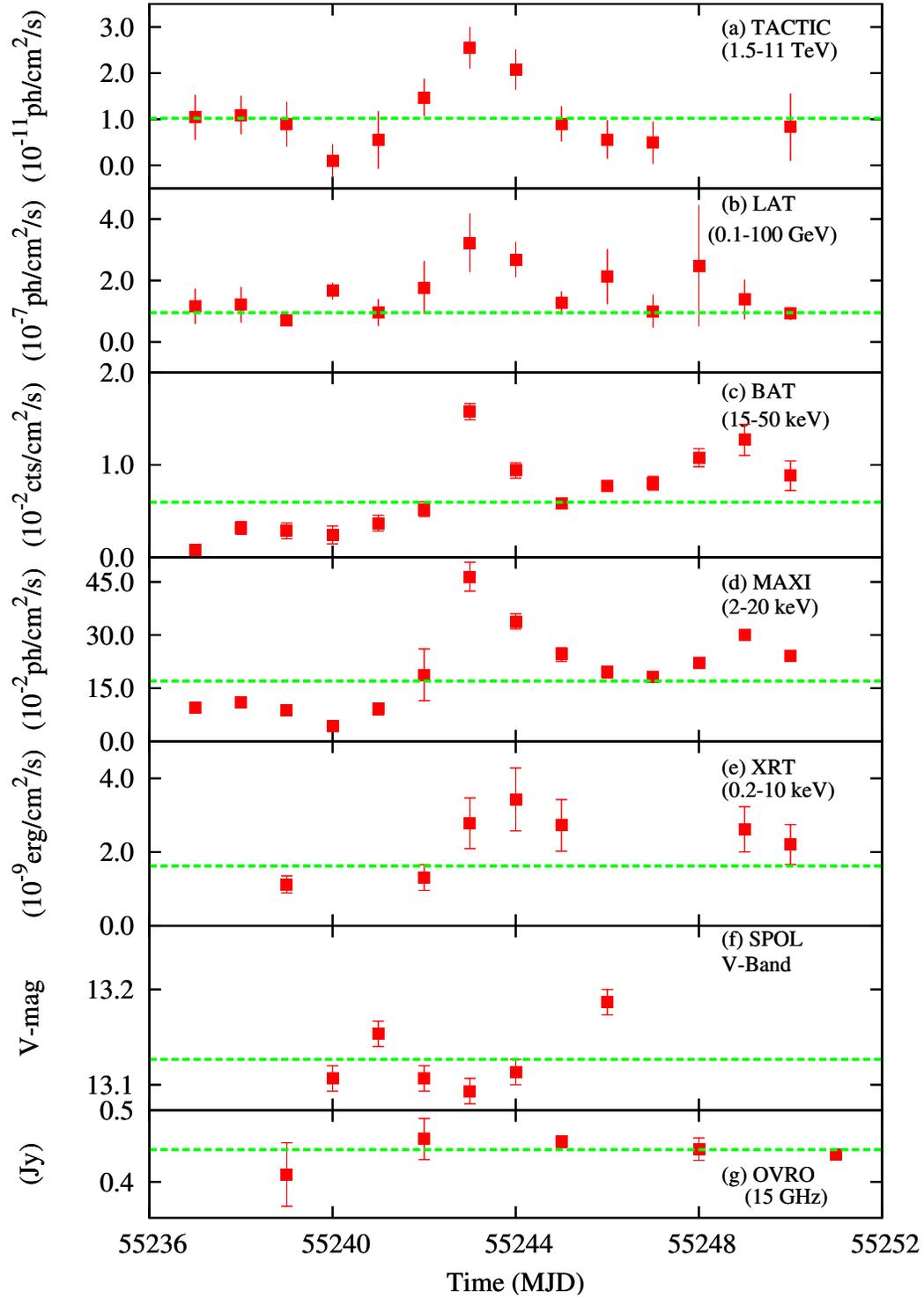}
\caption {Multi-wavelength light curves for Mrk 421 during February 10--23, 2010. The horizontal dotted lines represent 
the average emission during this period in each energy band.}
\end{center}
\end{figure}
\subsection{Variability and Correlations}
To quantify the flux variation in a given energy band, we deduce its fractional variability amplitude ($F_{var}$) given 
by [41],
\begin{equation}
	F_{var}=\frac{\left[S^{2}-\langle \sigma_{err}^{2}\rangle \right]^{\frac{1}{2}}}{\langle F\rangle} 
\end{equation}
where $\langle F\rangle$ is mean photon flux, \textit{S} is standard deviation  and $\sigma_{err}^{2}$ is  mean square error 
of N-flux points in the light curve. The error in fractional variability amplitude is then given by:
\begin{equation}
       \Delta 	F_{var}=\frac{1}{F_{var}} \sqrt{\frac{1}{2N}} \frac{S^2}{{\langle F \rangle}^2}
\end{equation}
The fractional variability amplitude as a function of mean observational energy of different instruments is shown in Fig. 3. 
Within the frame work of leptonic SSC model, the X--ray and $\gamma$--ray variabilities give information about the 
dynamics of relativistic electron population. From the figure, it is clear that the fractional variability amplitudes 
in various energy regimes are confined in the range 30--80$\%$. The fractional variability measured in X--rays and TeV 
$\gamma$--rays is apparently higher than that observed in other energy bands. This further suggests that these energies 
may be associated with the same population of relativistic electrons with faster cooling rates, favouring synchrotron and SSC 
origin of X-rays and TeV $\gamma$--rays. Whereas, the emission at other energies, including MeV-GeV, may be associated with low 
energy electrons with slower cooling rates. However, with the large error bars it is difficult to quantify the exact energy 
dependence of the fractional variability amplitude during the present flaring activity of the source. 
\begin{figure}
\begin{center}
\includegraphics*[width=0.8\textwidth,angle=-90]{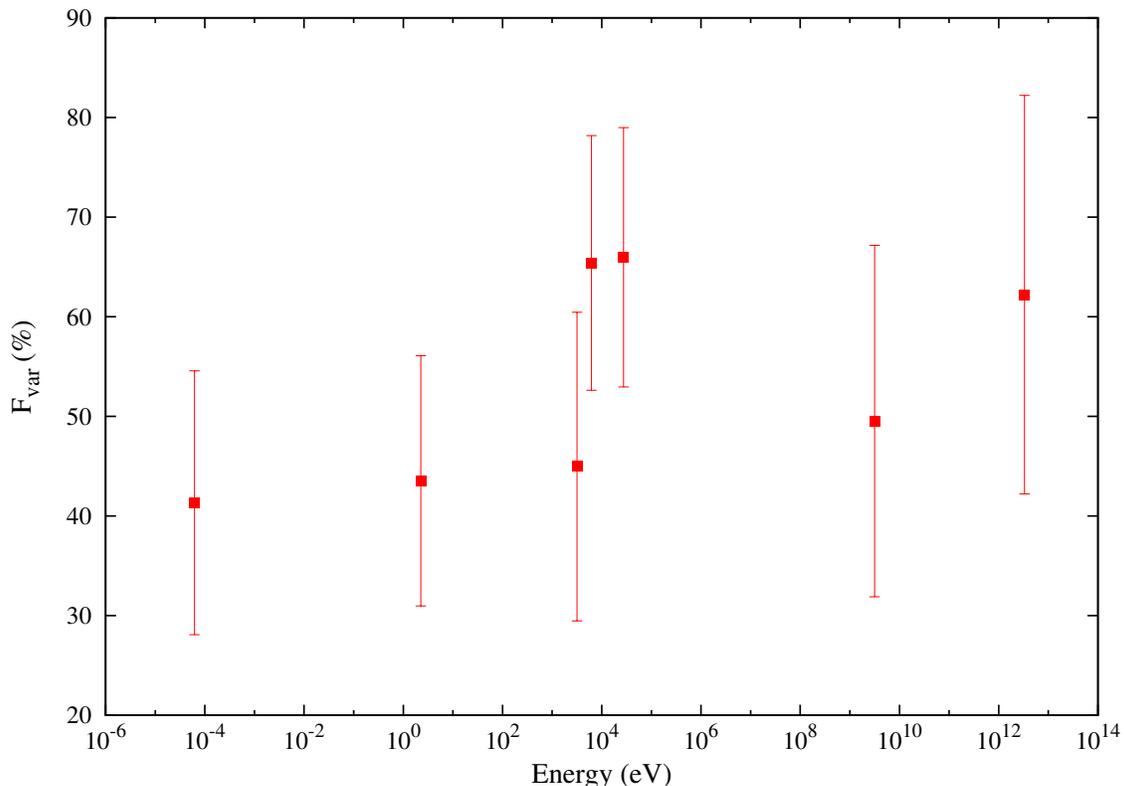}
\caption{Fractional Variability Amplitude in the multi-wavelength observations of Mrk 421 during February 10--23, 2010.}
\end{center}
\end{figure}
\par
We also compute the variability amplitude parameter ($A_{mp}$) introduced by [42] to characterize the percentage variation 
in each light curve. The variability amplitude parameter ($A_{mp}$) is defined as:
\begin{equation}
	A_{mp}=\frac{1}{\langle F \rangle} \sqrt{(F_{max}-F_{min})^2-2 \sigma^{2}} 
\end{equation}
where $F_{max}$ and $F_{min}$ are the maximum and minimum fluxes in each light curve and $\sigma$ is average measurement error in 
the light curves. The fractional variability amplitude ($F_{var}$) and variability amplitide parameter ($A_{mp}$) for 
MW observations obtained in the present study are given in Table 3. We further study the Pearson correlations between 
TeV and other simultaneous low energy emissions. The scatter plots for correlations with corresponding correlation coefficients 
are shown in Fig. 4(a-f). We observe that variations at X--ray energies observed by \emph{Swift}--XRT/BAT and MAXI are strongly 
correlated with TeV $\gamma$--rays, supporting our earlier inference that the same population of electrons is responsible for 
emission at these energies.
\begin{table}
\caption{Variability amplitudes for Mrk 421 multi-wavelength observations during February 10-23, 2010 (MJD 55237--55250).}
\vspace{0.3cm}
\begin{center}
\begin{tabular}{lccc}
\hline
Instruments  &Energy Band		&$F_{var}$ ($\%$)	&$A_{mp}$ ($\%$) \\ 
\hline
TACTIC	&1.5-11 TeV   		&62.2$\pm$20.0 		&209\\ 
LAT	&0.1-100 GeV		&49.5$\pm$17.6		&140\\
BAT	&15-50 keV  		&65.9$\pm$13.0		&198\\
MAXI	&2-20 keV   		&65.4$\pm$12.8		&193\\
XRT	&0.2-10 keV 		&44.9$\pm$15.4		&80\\	
SPOL	&V-band 		&43.5$\pm$12.5		&23\\
OVRO	&15 GHz 		&41.3$\pm$13.2		&7\\
\hline
\end{tabular}
\end{center}
\end{table}
\begin{figure}
\begin{center}
\includegraphics*[height=0.49\textwidth,angle=-90]{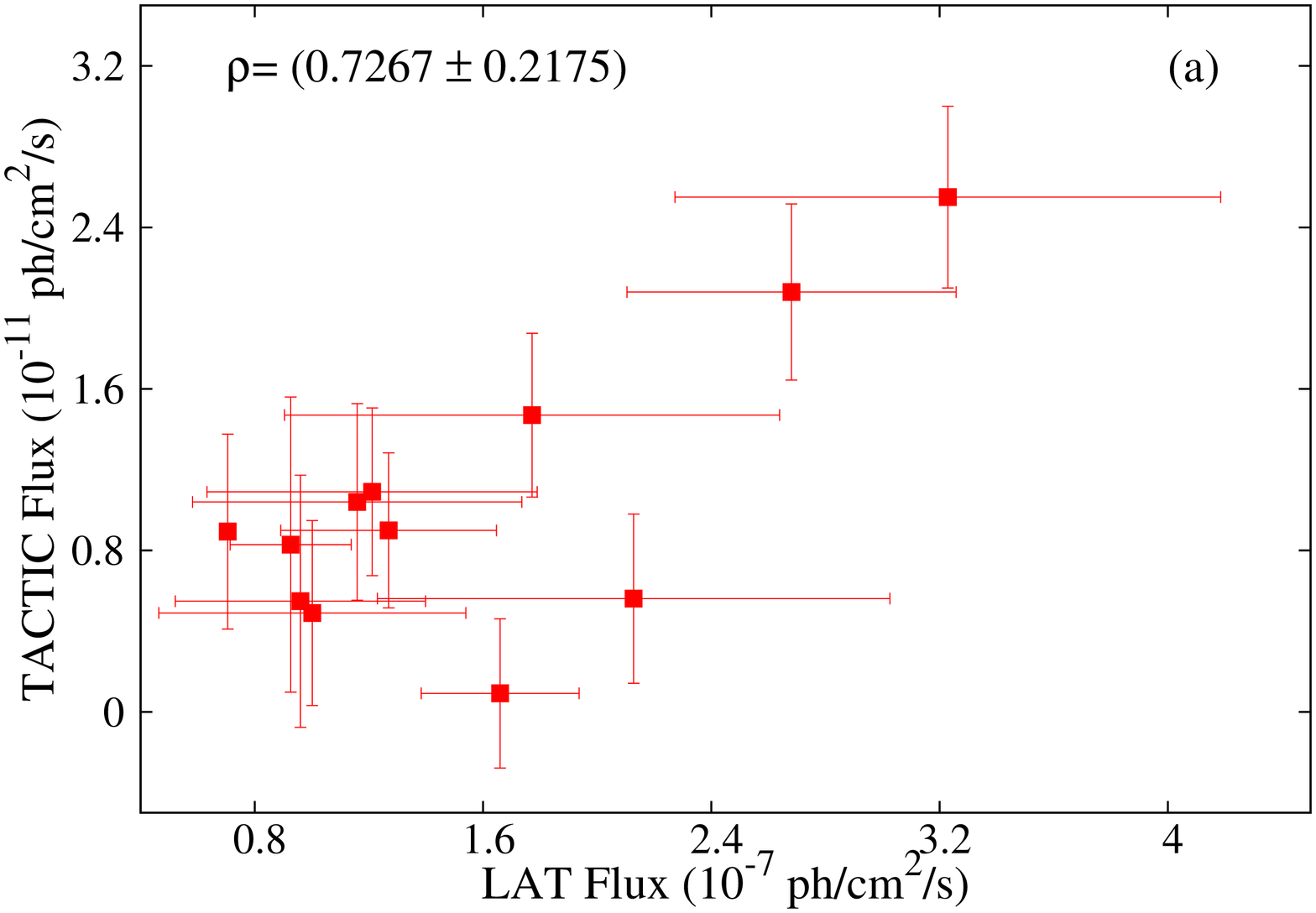}
\includegraphics*[height=0.49\textwidth,angle=-90]{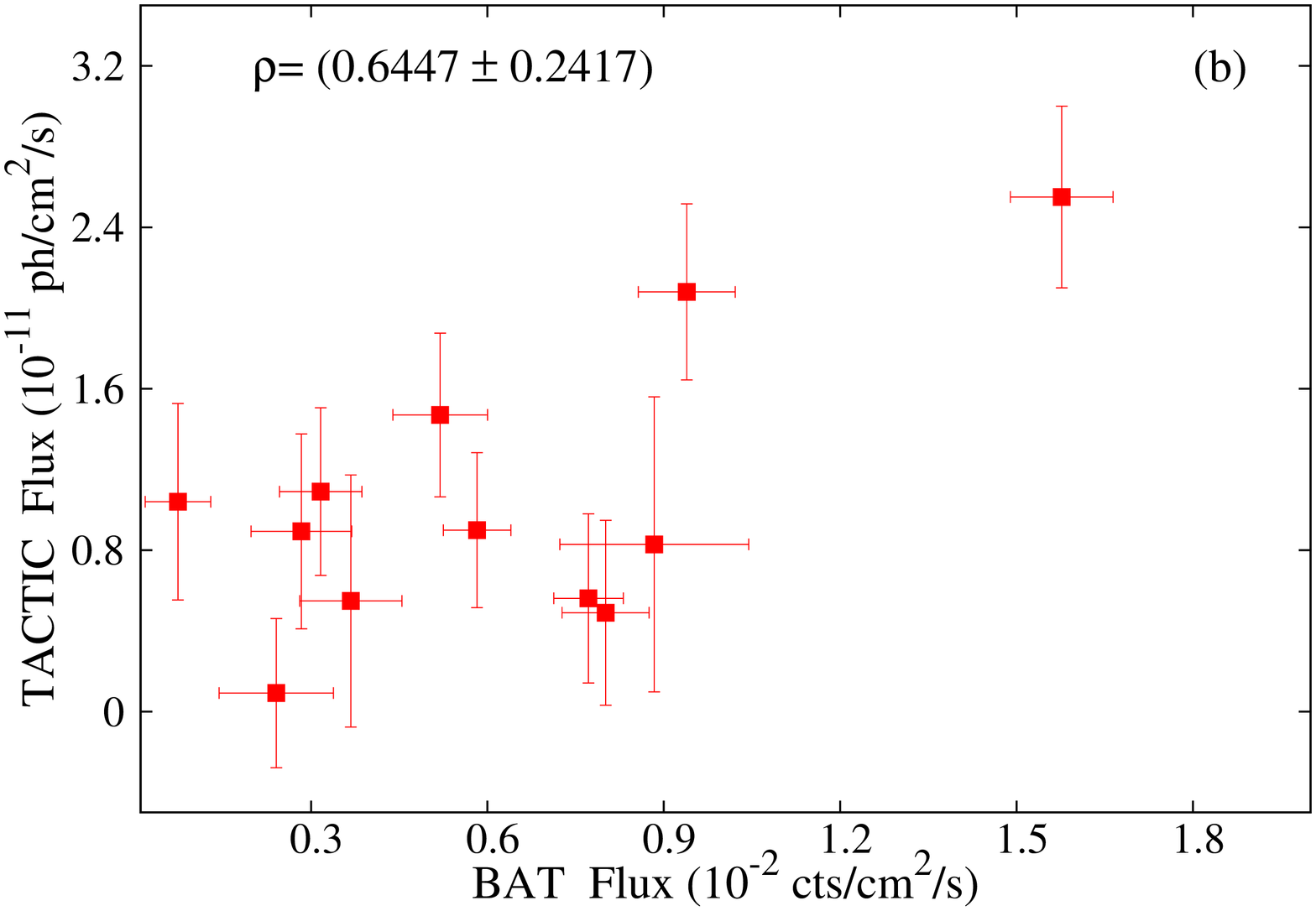}
\includegraphics*[height=0.49\textwidth,angle=-90]{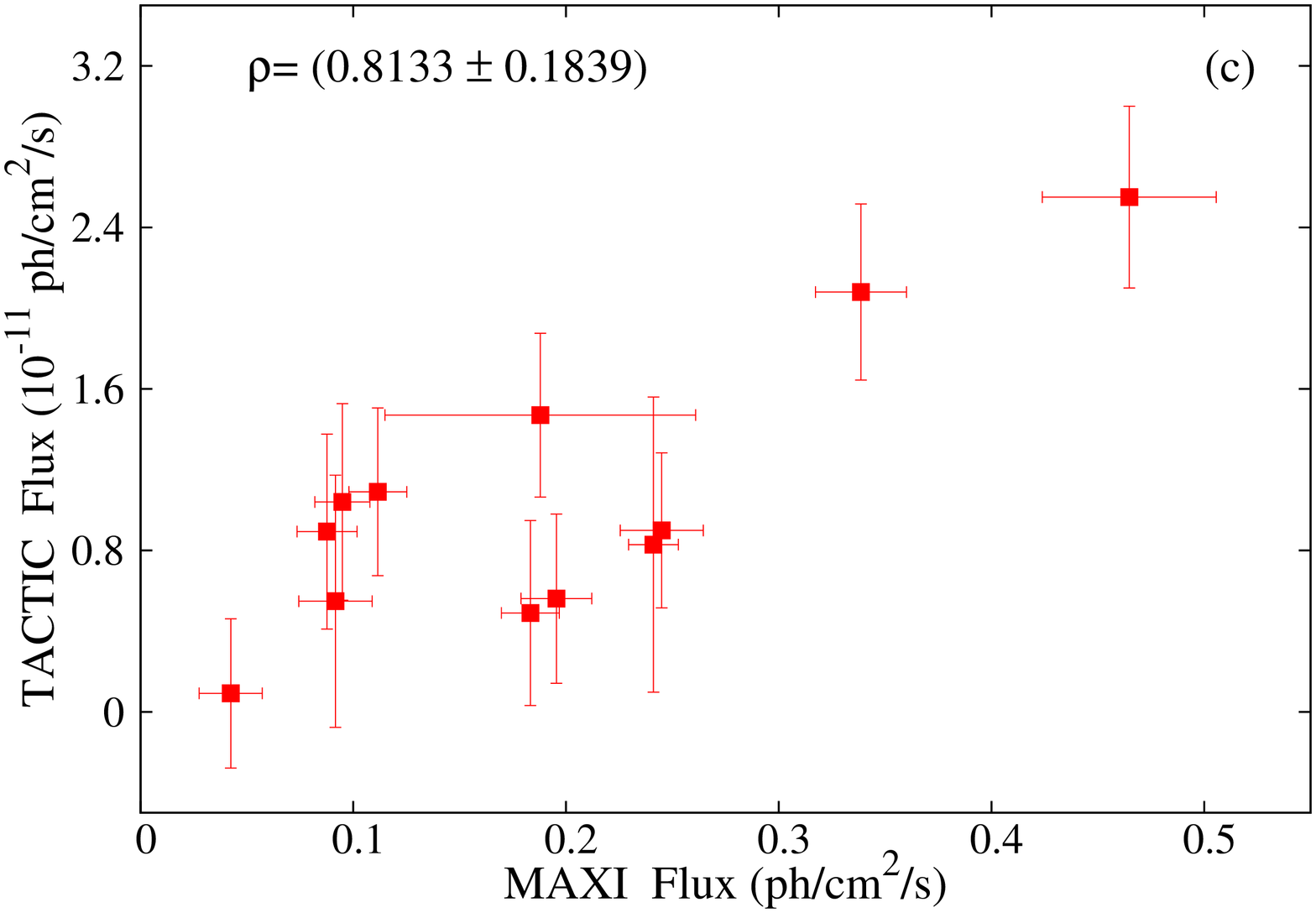}
\includegraphics*[height=0.50\textwidth,angle=-90]{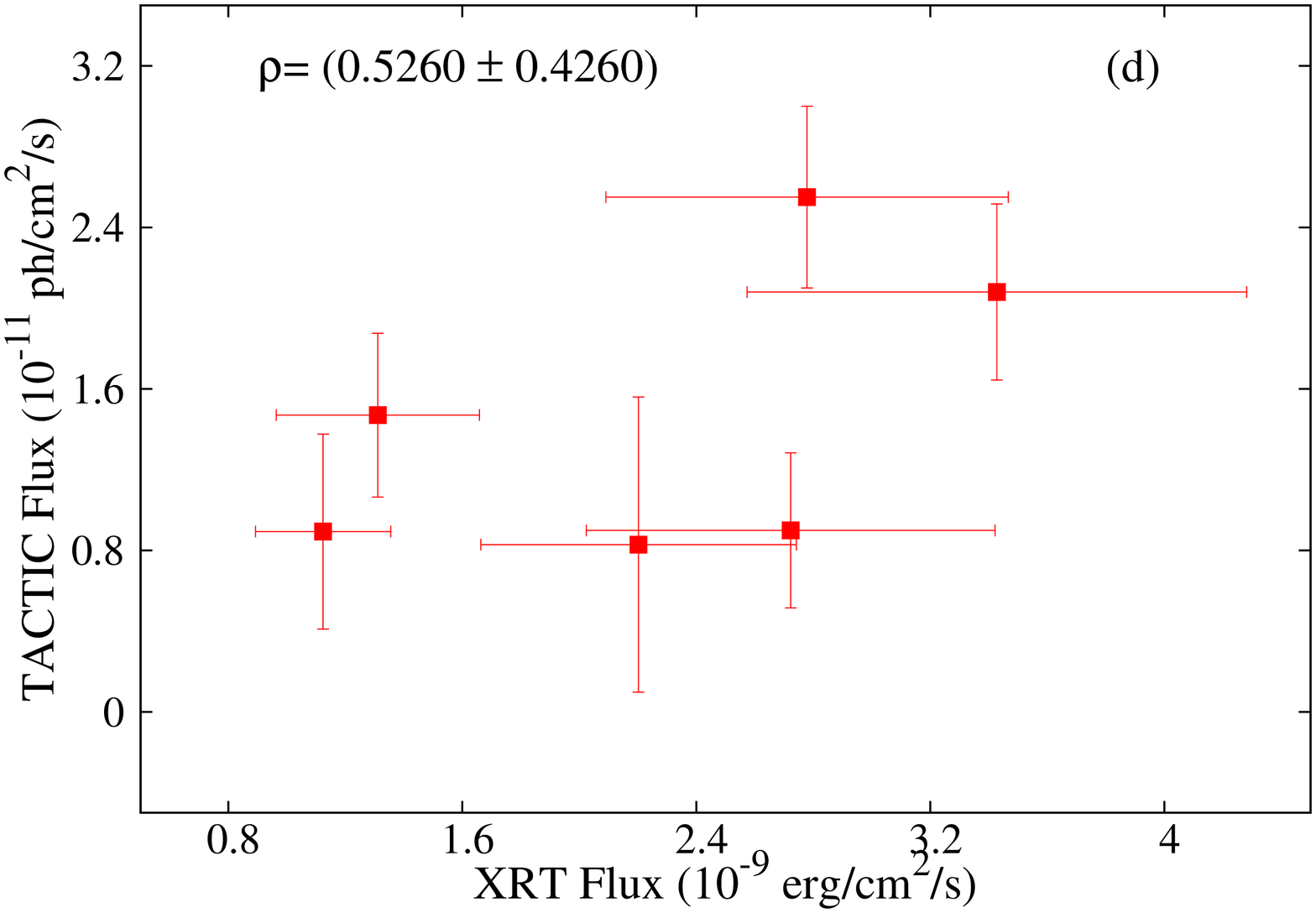}
\includegraphics*[height=0.49\textwidth,angle=-90]{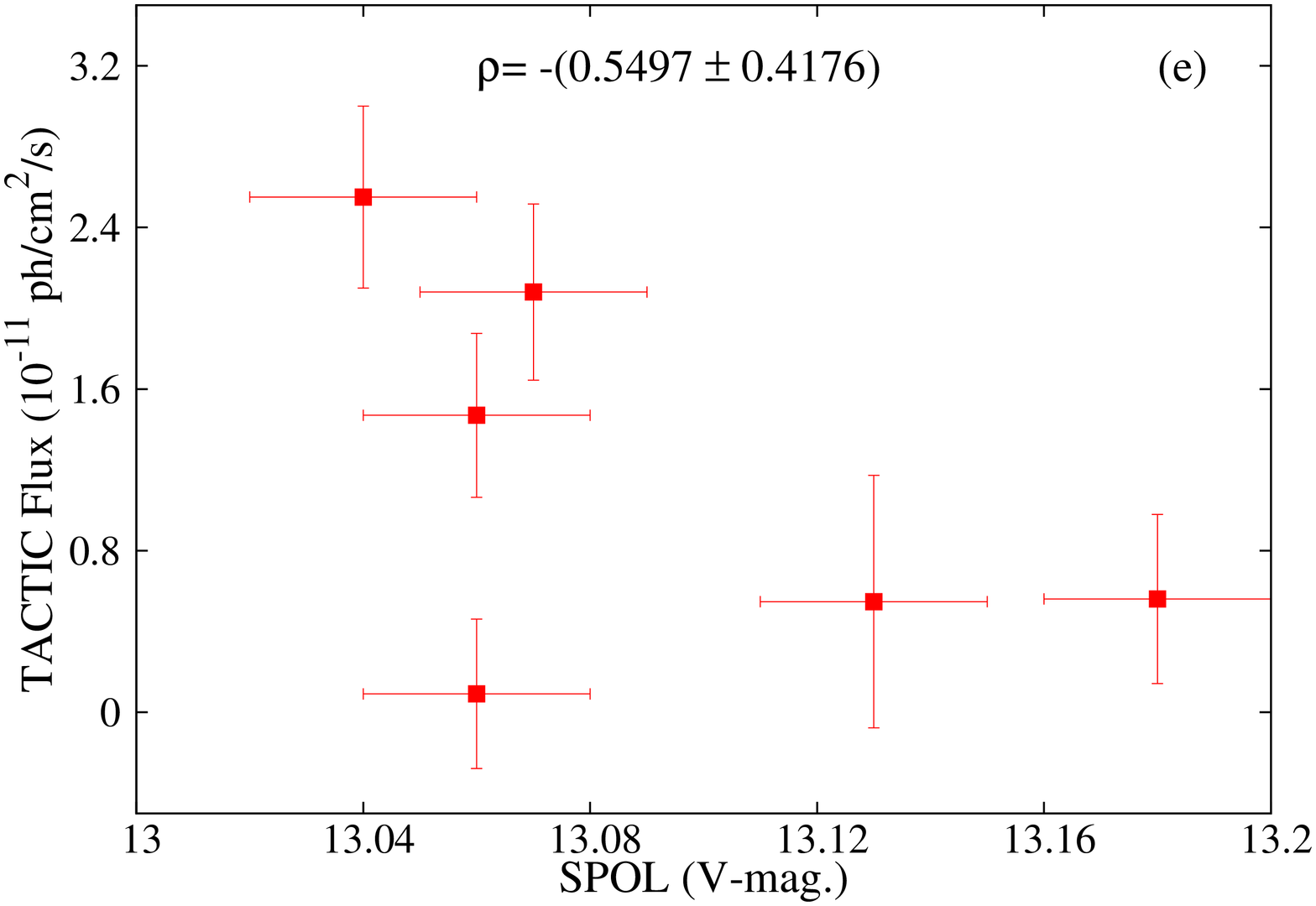}
\includegraphics*[height=0.49\textwidth,angle=-90]{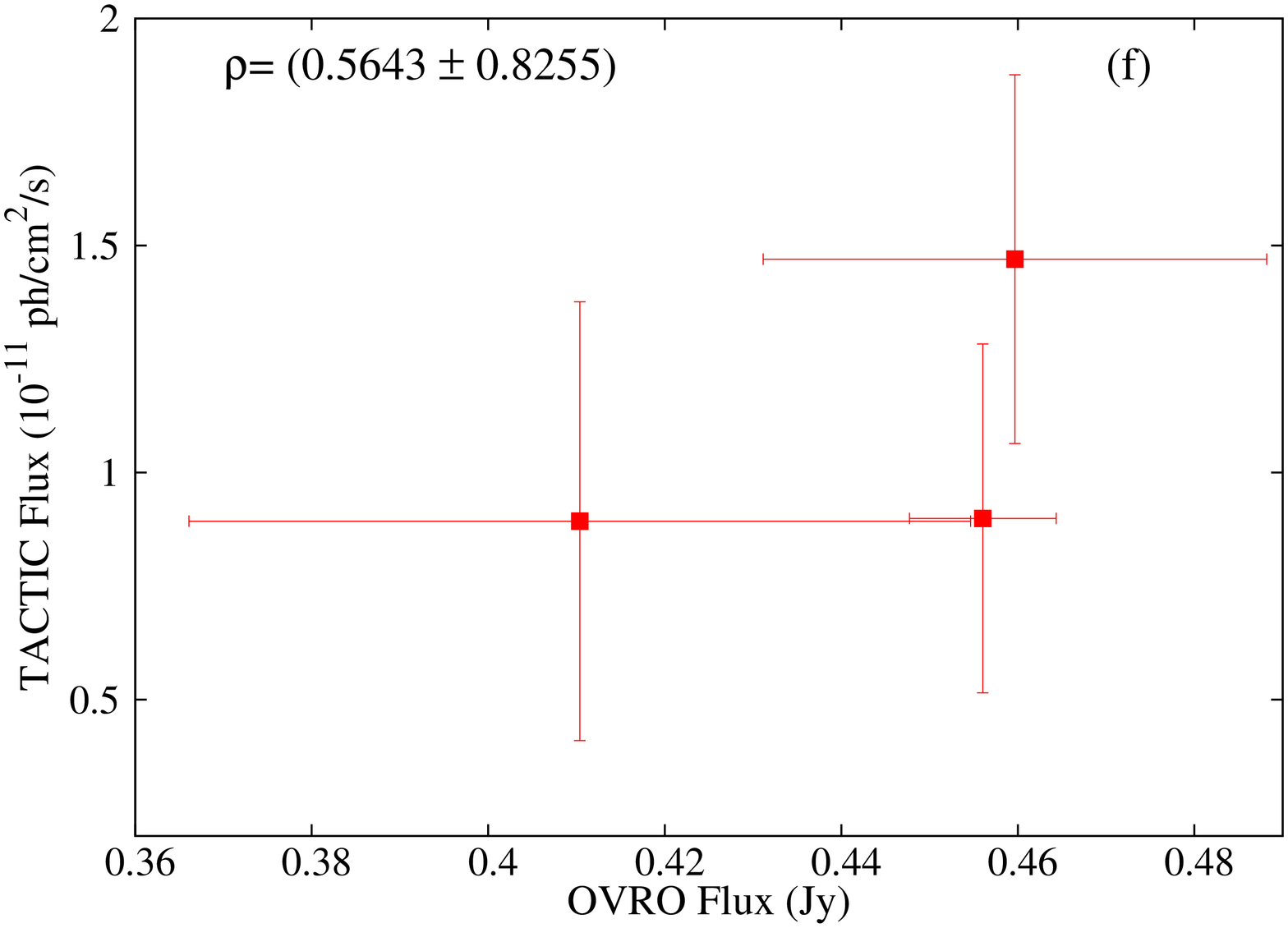}
\caption{Scatter plot for correlation between TeV $\gamma$-rays measured by TACTIC telescope and near simultaneous 
low energy observations.}
\end{center}
\end{figure} 
\section{Spectral analysis of flare on February 16, 2010}
\subsection{$TACTIC$ spectral analysis}
We use one day data comprising of $\sim$ 5 hours of observation on February 16, 2010 (MJD 55243) which corresponds to the flaring 
state of the source as seen with the $TACTIC$. This observation reveals an excess of 172$\pm$30 $\gamma$--ray like events with  
statistical significance of 5.92 $\sigma$. The corresponding observed differential energy spectrum of the source is presented in 
Fig. 5 and the respective flux points are given in Table 4. The differential energy spectrum of $\gamma$--rays from Mrk 421, in 
the energy range 1.5-11 TeV, as measured by $TACTIC$ on February 16, 2010 is described by a power law of the form :
\begin{equation}
 	\left(\frac{dN}{dE}\right)=(8.13 \pm 2.95) \times 10^{-11} \left(\frac{E}{1 TeV}\right)^{-2.60 \pm 0.35} 
						 	ph. cm^{-2}  s^{-1} TeV^{-1}
\end{equation}
In order to account for EBL absorption we have used two recent EBL models proposed by Franceschini et al. (2008) [43] 
and Dominguez et al. (2011) [44] to estimate the opacity of the Universe for TeV $\gamma$--rays from Mrk 421. The EBL 
corrected spectrum is again well described by a power law. The observed and intrinsic spectral parameters of the source 
are given in the Table 5. Since the intrinsic spectral indices corresponding to two EBL models are found to be similar, 
one can use either to model the SED of the source. If we attribute the TeV emission to inverse Compton (IC) emission from 
a power law distribution of electrons, the obtained intrinsic TeV spectral index corresponds to a particle spectral 
index $\sim$ 3.6.
\begin{figure}
\begin{center}
\includegraphics*[width=1.0\textwidth,angle=0]{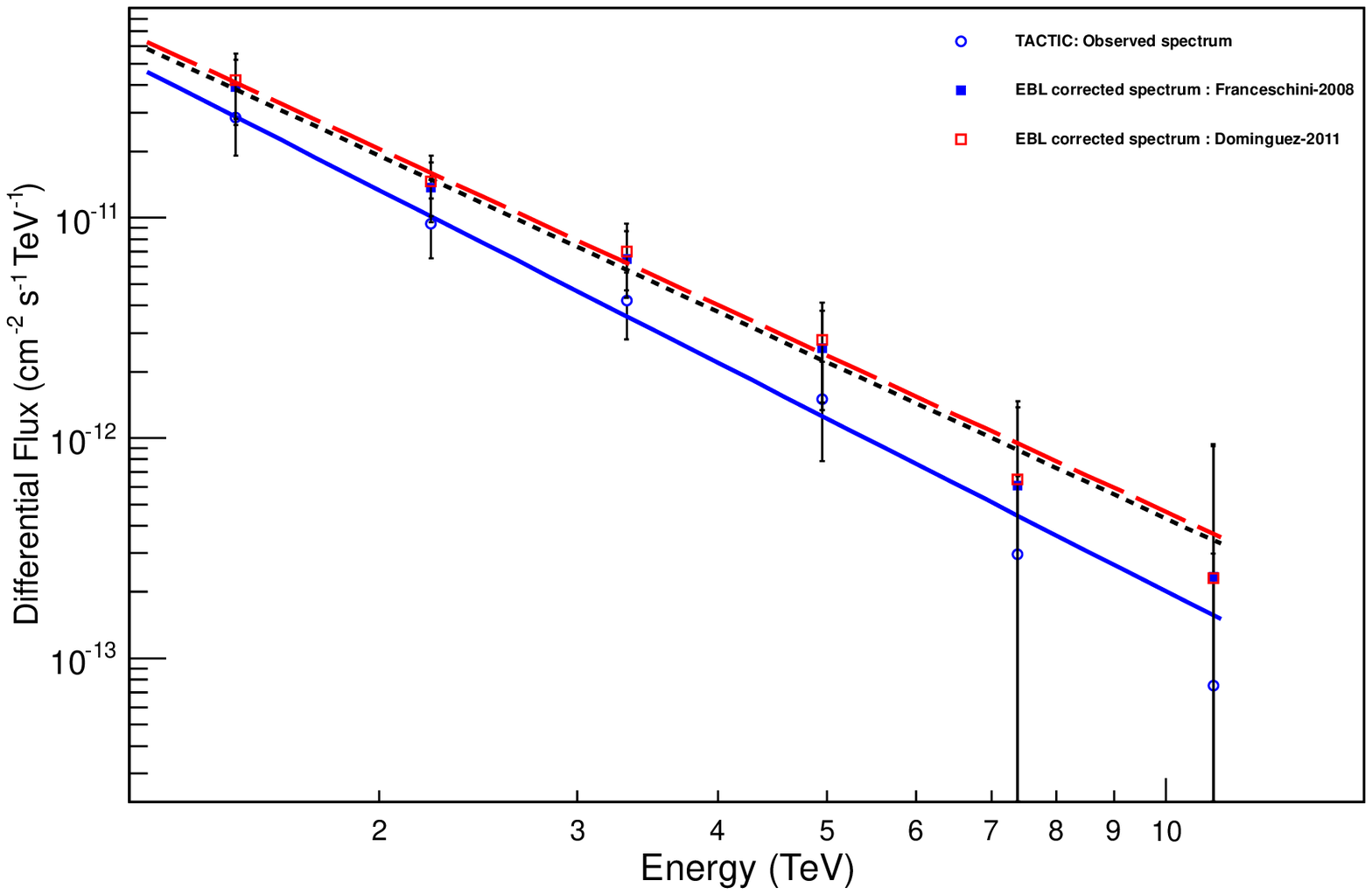}
\caption{Differential energy Spectrum of Mrk 421 observed with $TACTIC$ during the flare on  February 16, 2010. 
         De-absorbed source spectra for two EBL models are also shown.}
\end{center}
\end{figure}
\begin{table}
\caption{Differential energy spectrum flux points for Mrk 421 during flare on  February 16, 2010  measured with $TACTIC$. 
	Only statistical errors are  given.}
\label{tab:421sp_10}
\vspace{0.3cm}
\begin{center}
\begin{tabular}{lccc}
\hline
Energy      &Differential flux    		  &Statistical error in flux\\
(TeV)       &(photons cm$^{-2}$ s$^{-1}$TeV$^{-1}$) &(photons cm$^{-2}$ s$^{-1}$ TeV$^{-1}$) \\
\hline 
1.50        &2.84 $\times$10$^{-11}$              &9.34 $\times$10$^{-12}$\\
2.22        &9.36 $\times$10$^{-12}$              &2.84 $\times$10$^{-12}$\\
3.32        &4.21 $\times$10$^{-12}$              &1.41 $\times$10$^{-12}$\\ 
4.95        &1.50 $\times$10$^{-12}$              &7.15 $\times$10$^{-13}$\\
7.38        &2.96 $\times$10$^{-13}$              &3.75 $\times$10$^{-13}$\\
11.00       &0.75 $\times$10$^{-13}$              &2.24 $\times$10$^{-13}$\\
\hline
\end{tabular} 
\end{center}
\end{table}
\begin{table}
\caption{Paremeters of intrinsic differential energy spectrum of Mrk 421 fitted with power law ($f_0 E^{-\Gamma}$ where $f_0$ is 
flux normalization in $ph$ $cm^{-2} s^{-1} TeV^{-1}$ and $\Gamma$ is spectral index, $\Delta f_0 $ and $\Delta \Gamma$ are 
the corresponding uncertainties) for two EBL models.}
\vspace{0.3cm}
\begin{center}
\begin{tabular}{lccccc}
\hline
$\Gamma$ 	&$\Delta\Gamma$   &f$_{0}$   			&$\Delta$f$_{0}$   	      &Model\\ 
\hline
2.60		&0.35		  &8.13 $\times$ 10$^{-11}$	&2.95 $\times$ 10$^{-11}$     &	observed spectrum    \\  
2.35  		&0.38 		  &9.80  $\times$ 10$^{-11}$	&3.78 $\times$ 10$^{-11}$     &	EBL corrected [43] \\  
2.36 		&0.37		  &1.10 $\times$10$^{-10}$	&4.02 $\times$ 10$^{-11}$     &	EBL corrected [44]  \\  
\hline
\end{tabular}
\end{center}
\end{table}
\subsection{\emph{Fermi}--LAT spectral analysis}
The highest activity state of the source on Febrauary 16, 2010 was also observed by \emph{Fermi}--LAT in the energy range 
0.1--100 GeV with TS value 87 corresponding to statistical significance of 9.3$\sigma$. The LAT differential energy 
spectrum of the source during the flare has been obtained by dividing the LAT energy range into four energy bands: 0.1--1 GeV, 
1--3 GeV, 3--10 GeV and 10--100 GeV. The spectrum of Mrk 421 measured on February 16, 2010 by $LAT$ detector is described by 
a power law (d$\Phi$/dE=f$_0$ E$^{-p}$) with  f$_0$=(5.73$\pm$0.89)$\times$ 10$^{-8}$ cm$^{-2}$ s$^{-1}$ GeV$^{-1}$ 
and $p$= 1.72$\pm$ 0.13. Again, if we consider an IC origin of photons above 100 MeV from a power law distribution of electrons, 
this spectral index corresponds to a particle index $\sim$ 2.2. This particle index is considerably flatter than the one obtained 
from TeV spectral analysis and it cannot be associated with cooling effect, since their difference is not unity. Hence, the 
underlying particle distribution may be a broken power law probably resulting from multiple acceleration processes [45].
\section{Spectral modelling}
Motivated by the above temporal and spectral study, the time averaged SED of Mrk 421 during the flare on February 16, 2010 is studied 
using simple leptonic model involving synchrotron and SSC processes since this provides a simple explaination for MW emission from 
blazars [46]. The MW data are collected from optical, X--ray and VHE $\gamma$--rays with SPOL optical telescope, \emph{Swift}--XRT/BAT 
and $TACTIC$ observations respectively. The TeV flux points from $TACTIC$ are corrected for EBL absorption using model
proposed by Franceschini et al.(2008) [43]. To reproduce the broad band SED we adopt a model described in [47] where the 
emission region is assumed to be a spherical blob moving down the jet with bulk Lorentz factor $\Gamma$. The radius of blob 
$R$ is constrained by variability time scale ($t_{var}$) using the relation:
\begin{equation}
	R \approx\frac{c t_{var} \delta}{(1+z)} 
\end{equation}
where $\delta =[\Gamma(1-\beta cos \theta)]^{-1}$, is the Doppler factor with $\beta$ as the dimensionless bulk velocity and 
$\theta$ is angle between the jet axis and line of sight of the observer. Since blazar jet is aligned close to the line of sight, 
we can approximate $\delta=\Gamma$ corresponding to a viewing angle $\theta=cos^{-1} (\beta)$. Based on our temporal study, we 
consider $t_{var}$ $\sim$ 1 day to constrain the size of emitting region. The emission region is populated uniformly with a 
broken power law electron distribution described by:
\begin{equation}
	N(\gamma)d \gamma = K\left[\left(\frac{\gamma}{\gamma_b}\right)^{p1}+\left(\frac{\gamma}{\gamma_b}\right)^{p2}\right]^{-1} 
	d \gamma \quad ;\gamma_{min}< \gamma< \gamma_{max}
\end{equation}
where $K$ is the normalization, $\gamma_b m_e c^2$ is the break energy with $m_e$ as the electron rest mass, $p1$ and $p2$ are the 
power law indices before and after the break energy $\gamma_b m_ec^2$. $\gamma_{min}m_ec^2$ and $\gamma_{max}m_ec^2$ are the 
minimum and the maximum electron energies of the distribution. The particles lose energy through synchrotron emission in a magnetic 
field $B$ and synchrotron self Compton emission. The magnetic field energy density is considered to be in equi-partition with 
that of electrons,
\begin{equation}
	U_B = m_e c^2 \int\limits_{\gamma_{min}}^{\gamma_{max}} \gamma N(\gamma)  d\gamma = U_e
\end{equation}
where $U_B = B_{eq}^2/8\pi$ is the magnetic field energy density and $U_e$ is the particle energy density. Due to relativistic 
bulk motion, the radiation from the emission region is Doppler boosted by $\delta^3$. The main model parameters are constrained 
by the results obtained through our temporal and spectral studies and the resultant SED along with observed fluxes are shown in 
Fig 6. The parameter values estimated in the present work have been summarized in Table 6 and are consistent with the values 
reported in the literature [48,49]. Due to large uncertainties in flux points, the LAT data are represented as butterfly plot 
in the figure. The VHE flux points reported by $VERITAS$ and $HESS$ telescopes are also shown in the figure and have been corrected 
for EBL absorption [43]. 
\begin{figure}
\begin{center}
\includegraphics*[width=0.7\textwidth,angle=-90]{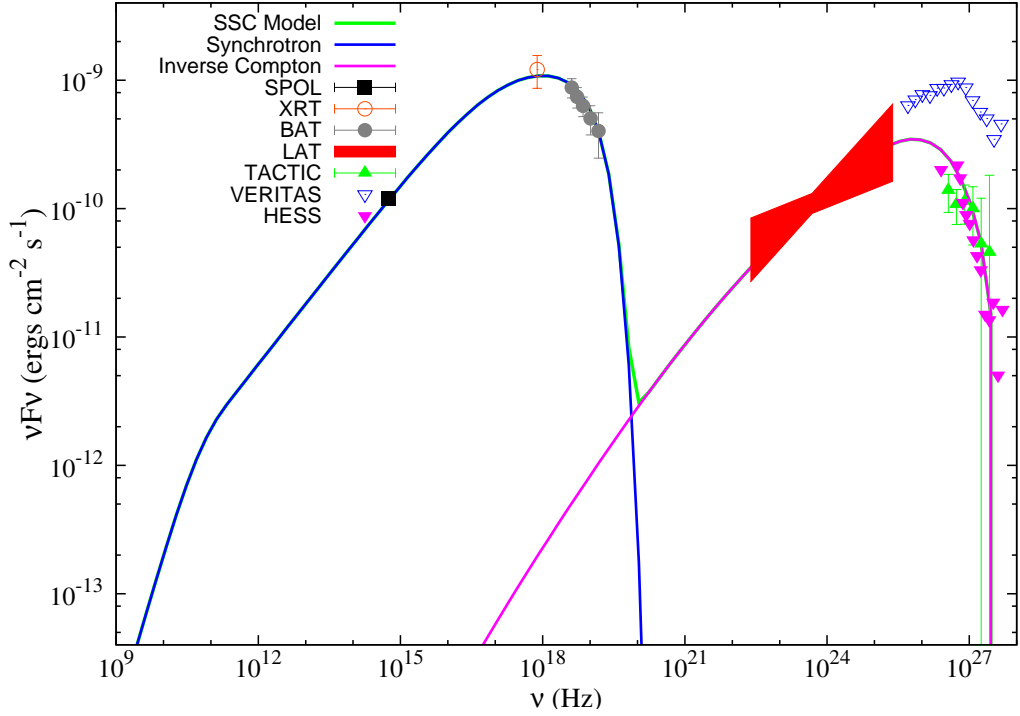}
\caption{Spectral energy distribution of Mrk 421 using one-zone SSC model during flare on February 16, 2010 (MJD 55243).
         $TACTIC$ flux points are corrected with Franceschini-2008 EBL model [23]. The VHE flux points obtained from  
        $VERITAS$ [25] and $HESS$ [26] telescopes have been corrected for EBL absorption [23]. The $VERITAS$ flux points 
        correspond to the observation on February 17, 2010 (MJD 55244) whereas $HESS$ points are time averaged fluxes for 
        observations during February 17-20, 2010 (MJD 55244-55246).}
\end{center}
\end{figure} 
\begin{table}
\caption{Optimized source parameters and properties of Mrk 421 obtained by fitting one day MW data during flare on February 16, 
2010 using simple one zone SSC model.}
\vspace{0.3cm}
\label{Table:SSC}
\begin{center}
\begin{tabular}{lcc}
\hline
Parameter           		& Value \\
\hline
Blob Radius 	    		& $2.37\times10^{16}$cm\\        
Bulk Lorentz factor 	        & 14\\
Break energy of particle distribution &168 GeV\\    
Particle energy density 	      & $2.42\times10^{-3}$erg/$cm^{3}$\\   
Power law index before break          &2.22\\
Power law index after break	      &3.80\\
Magnetic field 			      &0.36 G\\
Jet Power                             &$10^{44}$ erg/s\\
Radiated Power			      &$10^{42}$ erg/s\\
\hline
\end{tabular}
\end{center}
\end{table}
\section{Discussion and Conclusion} 
In the present work we have used TeV observations with $TACTIC$  together with other MW data on Mrk 421, to study the variability 
and spectral properties of the source during its high state of activity. Apart from analyzing the $TACTIC$,\emph{Fermi}--LAT and 
\emph{Swift}--XRT data during Febraury 10--23, 2010 (MJD 55237--55250), we have also used simultaneous archival data for 
\emph{Swift}--BAT, MAXI, optical (V-band) and radio (15 GHz) observations for our study. The highest flaring activity from 
the source in TeV energy band, as measured by $TACTIC$ was detected on February 16, 2010 (MJD 55243) and  an enhanced 
activity was also observed in  HE $\gamma$--rays from \emph{Fermi}/LAT and X--rays. 
\par 
The varaiability studies in various energy bands (Fig.3) suggest the emission to arise from the jet. Similar features observed in 
X-ray and TeV flare also support synchrotron and SSC origin of these emissions. The data statistics are not good enough to assert 
these interpretations strongly. Especially, the double flare seen in X--ray energy band, reflect the amount of complexity involved 
in these emission processes. Modelling such features may demand detailed study involving various acceleration mechanisms driving 
blazar flares; however, this is beyond the scope of the present study. Recently, Dahai Yan et al. (2013) have investigated the 
electron energy distributions and the acceleration mechanisms in the jet of Mrk 421 by fitting the SED in different active states 
under the framework of single-zone SSC model [50]. They conclude that the shock acceleration is dominant in low activity state, while 
stochastic turbulence acceleration is dominant in flaring state. Whereas, Mastichiadis et al. (2013) have studied the origin of 
$\gamma$--ray emission in blazars within the context of the lepto-hadronic single zone model [51]. They find that $\gamma$--ray 
emission can be attributed to synchrotron radiation either from protons or from secondary leptons produced via photohadronic 
process. These possibilities imply differences in the X--ray and $\gamma$--ray variability signatures. 
\par 
The time averaged broad band SED of Mrk 421 during the flare can be well reproduced  under the framework of simple one zone 
SSC model. The resulting parameters from the SED modeling are consistent with the source parameters used in SSC model 
reported in the literature [48,49]. From these parameters  we estimate the kinetic power of the jet ($P_{jet}$) by assuming 
that the emission region is also populated with cold protons equal in number as that of the non thermal electrons. 
The power of the jet can then be approximated as [52]
\begin{equation}
	P_{jet} \approx \pi R^{2} \Gamma^{2} \beta c (U_{p}+U_{B}+U_{e})
\end{equation}
where $U_{p}$, $U_{B}$ and $U_{e}$ are cold proton energy density, magnetic field energy density and electron energy density in the 
rest frame of emission region respectively. The obtained jet power ($P_{jet}$=$10^{44}$ erg/s) is consistent  with the one generally 
assumed for blazars and is much larger than the power released in the form of radiation ($P_{rad}$=$10^{42}$ erg/s). Hence at blazar 
zone only a small fraction of the jet kinetic energy is utilized in radiation and most of the energy is spent in driving the jet 
upto $Mpc$ scale. Eventhough, the obtained model parameters are consistent with the general accepted values for Mrk 421, they differ 
considerably from the one reported by [20] during the same flaring episode. We note that the primary reason for this difference 
is due to the constraint introduced through equi-partition condition (equation 8). Since equi-partition assures that the system is under 
minimum energy state (stable) [53], the parameters obtained in the present work may be the more probable ones. However, it is also 
to be noted that many blazars do not satisfy the equi-partition condition during a flare [54]. Most of these uncertanties regarding 
the emission models and underlying parameters can be resolved  by detailed modelling of blazar light curves using complex algorithms. 
However such work would involve large number of parameters and future simultaneous MW observations of blazars during flare can 
be used to estimate/constrain these parameters.   
      
\section*{Acknowledgements} 
We would like to acknowledge the excellent team work of our colleagues in the Division for their contribution to 
the simulation, observation, data analysis and instrumentation aspects of $TACTIC$ at Mt. Abu observatory. We 
acknowledge Ms. Atreyee Sinha, TIFR, Mumbai for providing \emph{Swift}/BAT spectral data used in 
SED modelling of the source. We acknowledge the use of public data obtained through \textit{Fermi} Science Support 
Center (FSSC) provided by NASA. This work made use of data supplied by the UK Swift Science Data Centre at the 
University of Leicester. This research has made use of the MAXI data, provided by RIKEN, JAXA and the MAXI team. 
Data from the Steward Observatory spectropolarimetric monitoring project were used. This program is supported by 
\emph{Fermi} Guest Investigator grants NNX08AW56G, NNX09AU10G and NNX12AO93G. Radio data at 15 Ghz is used from 
OVRO 40 M Telescope and this \emph{Fermi} blazar monitoring program is supported by NASA under award NNX08AW31G, 
and by the NSF under 0808050. We would like to thank the anonymous referee for his/her valuable comments and 
suggestions which have helped us in improving the quality of the paper.

\end{document}